# The promise of behavioral tracking systems for advancing primate animal welfare

**Brenna Knaebe, Claudia C. Weiss, Jan Zimmermann, and Benjamin Y. Hayden**

**Affiliations**
>1. Department of Neuroscience, Center for Magnetic Resonance Research, Department of
Biomedical Engineering
>University of Minnesota, Minneapolis MN 55455

**\*Corresponding author**
>Brenna Knaebe
>Department of Neuroscience
>Center for Magnetic Resonance Research
>University of Minnesota
>Minneapolis, MN, 55455
>email: knaeb046@umn.edu

**SIMPLE SUMMARY**

Computerized tracking systems for primates and other animals are one of the great inventions of the 21st century. These systems have already revolutionized the study of primatology, psychology, neuroscience, and biomedicine. Less discussed is that they also promise to greatly enhance animal welfare. Their potential benefits include identifying and reducing pain, suffering, and distress in captive populations, improving laboratory animal welfare and applying our understanding of animal behavior to increase the "natural" behaviors in captive and wild populations, especially those under threat. We are optimistic that these changes will greatly increase welfare in primates, including those in laboratories, zoos, primate centers, and in the wild.

**ABSTRACT**

Recent years have witnessed major advances in the ability of computerized systems to track the positions of animals as they move through large and unconstrained environments. These systems have so far been a great boon in the fields of primatology, psychology, neuroscience, and biomedicine. Here, we discuss the promise of these technologies for animal welfare. Their potential benefits include identifying and reducing pain, suffering, and distress in captive populations, improving laboratory animal welfare within the context of the three Rs of animal research (reduction, refinement, and replacement), and applying our understanding of animal behavior to increase the "natural" behaviors in captive and wild populations facing human impact challenges. We note that these benefits are often incidental to the designed purpose of these tracking systems, a reflection of the fact that animal welfare is not inimical to research progress, but instead, that the aligned interests between basic research and welfare hold great promise for improvements to animal well-being.

**INTRODUCTION**

Non-human primates - monkeys and non-human apes - are found in zoos and research settings in large numbers, and are a crucial animal model in the biomedical sciences (Rudebeck et al., 2019; Buffalo et al., 2019; Gray and Barnes, 2019; Picaud et al., 2019; Roberts and Clarke et al., 2019; Bernardi and Salzman, 2019). Captive non-human primates are uniquely susceptible to welfare challenges for several reasons. First, their social nature, their high level of intelligence, and the fact that they are not domesticated animals (Hau & Schapiro 2007). Wild populations also face their own set of welfare challenges, caused by encroachment of humans. These include problems caused by habitat loss, bushmeat hunting, the illegal pet trade, climate change, and anthroponotic diseases (Estrada et al 2017). Because of these welfare challenges, there is a growing ongoing interest in improving welfare for non-human primates. The goal of improving welfare is primarily a moral concern. However, secondary motivations include the benefits that improved welfare offers for improving the rigor and reliability of resulting scientific outcomes (Graham & Prescott 2015).

Recent years have witnessed great advances in technology that allows for the tracking of animals, including primates (Mathis & Mathis 2020; Mathis et al 2020; Pereira et al 2020). We use the term "behavioral imaging" to refer to these technologies (Hayden et al., 2021). Behavioral imaging has attracted a great deal of interest for its potential applications to neuroscience, psychology, zoology, and ethology, among other fields. However, the potential benefits of these technologies for primate welfare have not been well explored. Here we propose that, in addition to its other benefits, the widespread adoption of behavioral imaging will have salutary effects on animal welfare. Specifically, in the current paper we present three main domains where behavioral imaging can improve the welfare of non-human primates: (1) identifying and reducing pain, suffering, and distress in captive populations; (2) improving laboratory animal welfare within the context of the three Rs of animal research (reduction, refinement, and replacement); and (3) using applied animal behavior to promote "natural" behaviors in captive and wild populations facing human impact challenges.

**What is behavioral imaging?**

In the past, those interested in understanding the behavior of primates in detail had one option - describing in words what they could see with their own eyes. Technological advances have changed this. Now, we can use digital video cameras with software to monitor the positions of animals' bodies in space ("pose tracking"). Sophisticated analysis systems can then process this information about pose to determine the behavior the primate was performing (walking, climbing, grooming, etc) at every moment ("behavioral identification") in a continuous fashion. We use the term "behavioral imaging" to refer to both of these methods together. Here, as a prelude to exploring their applications to welfare, we briefly review these methods. For a longer review on behavioral imaging in primates, see Hayden et al. (2021).

***Pose tracking:*** Modern behavioral imaging relies on several major technologies, especially affordable high quality digital video cameras, image processing software, storage, and deep-learning techniques for analysis of data. These technologies allow for the tracking of primates with high spatial and temporal resolution, often from multiple vantage points, often for long periods of time. These systems record a digital impression of the scene in front of them and, following extensive training, can identify movement of humans and other animals in the scene. Fundamentally, these systems are based on detailed annotated training datasets that give example scenes and pose annotated animals. It is the need for such training sets, rather than technological inadequacies, that tends to serve as the major barrier to progress in these systems. The resulting systems can follow the positions of a few or several landmarks with high spatial and temporal precision. Multi-camera systems can readily provide depth information (the position of landmarks within the three dimensional scene), while single-camera systems can typically only provide information about the positions of landmarks in the frame. Some recent advances allow for the estimation of depth information from single-view monitoring (called "lifting") in some circumstances (Mehrizi et al 2018; Tome et al 2017; Zhou et al 2017).

***Behavioral identification:*** The detection of pose (the positions of major body landmarks in the scene) is a precursor to the identification of behavior (the categorical identity of what the animal is doing). Some behaviors are identifiable solely from pose - consider, for example, the distinction between walking and sitting. Others require contextual information. For example, a monkey may sit in the same position when it is foraging for insects to eat and when grooming its relatives. As a consequence, it is typically non-trivial to identify behaviors based on even perfect reconstructions of pose. Fortunately, recent research has begun to develop the ability to identify specific behaviors (Bialek, 2022; Marshall et al., 2021; Wiltschko et al., 2020; Vogt, 2018; Bohnslav et al., 2021; Dunn et al., 2021; Berman, 2018; Datta et al., 2019).

Together, these systems constitute tools for behavioral imaging. The fact that these systems do not require human supervision makes them orders of magnitude less expensive than having humans "in the loop". Their ability to process large volumes of data makes it possible to detect rare and subtle behaviors, and changes in behavior over time. Their ability to combine images from multiple views makes it possible to do complex imaging that humans cannot. The fact that they can classify behavior using unsupervised methods removes the inherent bias of subjective human annotation of behavior. For these reasons, these techniques have engendered a great deal of optimism for scientists interested in animal behavior and animal welfare (AbdulJabber et al., 2022; Costa-Periera et al., 2022; Tuia et al., 2022).

## BEHAVIORAL IMAGING CAN HELP ASSESS PAIN AND DISTRESS WHILE IMPROVING WELFARE

The first step in the improvement of welfare is to measure it. However, measuring welfare is a surprisingly difficult problem (Carstens & Moberg 2000). Any animal welfare guidelines will stress the importance of the reduction of pain and distress, but the literature often does not provide clear indications of how to detect them (Broom 1991). Indeed, it often takes a large degree of experience with animals for a veterinarian or caretaker to gain the intuition to detect pain or distress. Even then, the time, manpower, and expertise required to monitor animals for signs of pain or distress can be prohibitively costly.

**Current Challenges of Assessing Pain and Distress**

Researchers conducting biomedical studies on animals are responsible for reducing potential discomfort and promptly treating any pain that may arise (Coleman 2011). Nonhuman primates are of particular concern because of their high intelligence and their tendency to hide pain from observers. Indeed, macaques are notorious for masking clinical signs of illness and injury, possibly to conceal impairments from social group members and/or due to their evolutionary status as prey animals (Lefebvre & Carli, 1985; Gaither et al. 2014).

In order to detect pain, researchers must monitor physiological and/or behavioral measures that may indicate pain. Physiological responses are not particularly effective measurements of pain, as these responses are often delayed (e.g. cortisol measurements in urine); are potentially invasive and stressful in and of themselves (e.g. sampling blood) (Carstens & Moberg 2000); and single measures can indicate either negative or positive states, such as how changes in heart rate and cortisol/corticosterone levels from emotional arousal can indicate either fear or pleasure (Prescott & Lidster 2017). Poor health responses, while vital to animal welfare, are more likely to indicate extreme issues with welfare rather than more subtle issues; while pathological indicators of distress such as disease, self-injurious behaviors, or death will become apparent given enough time, it is unethical to wait for these outcomes to appear.

Because of the issues with physiological detectors, behavioral observations are the superior tool in gaining insight into an animal's state of welfare, especially early on. There are known indicators of pain in nonhuman primates, such as impaired locomotion, favoring limbs, over-grooming, etc. But because of their tendency to mask discomfort, nonhuman primates have a strong "observer effect" where they hide these indicators in the presence of an active observer. Research has found that nonhuman primates will resist exhibiting signs of pain and illness (such as hunched with head down, lying down, and dropping food from mouth) when an observer is actively attending to them, even when in moderate to severe states of pain (Gaither et al. 2014). Despite being more feasible and potentially more reliable than physiological measures, the observer effect makes it particularly challenging for researchers and caretakers to detect signs of discomfort in nonhuman primates. Because of the effect of direct observation on displays of pain and distress, the Institute for Laboratory Animal Research (ILAR) recommends that NHPs should be visually assessed from a distance. However, this approach does not eliminate observer effects and some behaviors may be subtle or may worsen over time such that an indirect observer will not notice. Video recording can replace behavioral assessments, as cameras will not trigger observer effects. However, these recordings must be annotated by a trained observer to determine which behaviors are abnormal for an individual and will cause a delay in treatment response. Furthermore, a veterinarian or highly trained staff member must be familiar with normal behaviors at both the species and individual level to ensure proper assessment.

As with pain, it can be challenging to recognize stress in animals. Stress is generally defined as any biological response to perceived threats in the animal's environment (National Research Council 2010). Stress is not always negative; in small doses, stress responses can help an animal run away from predators, promote social bonding, and/or create resilience to future stressors (Young et al. 2014; Meyer & Hamel 2014). However, when stressors result in negative pathological responses, the state of the animal becomes one of distress (Carstens & Moberg 2000). When a stressor is chronic, severe, and/or a culmination of several threats, the biological cost of reacting to the stressor may disrupt normal biological processes. This can eventually lead to pathological responses such as behavioral abnormalities, loss of reproduction, or growth abnormalities, to name a few (Maestripieri & Hoffman 2011).

Currently, one of the best ways to behaviorally measure stress and distress is by observing stereotypies (Dantzer 1991; Carstens & Moberg 2000; Fam, Tan, & Waitt 2012). Stereotypies are defined as repeated actions performed as a coping response to stress that have no obvious function to the animal (Crowell-Davis 2007). Stereotypies are assumed to originate as coping responses to environment-induced stress, such as a limited ability to perform species-specific natural behaviors or inadequate social and tactile enrichment (Fam, Tan, & Waitt 2012). The actions are not initially pathological, but eventually become a learned behavior tool used by the animal to cope with stressful situations. Stereotypical behaviors then integrate into the animal's behavioral repertoire to a degree that disrupts their normal activities. As a consequence, stereotypies can serve as an index of distress and can indicate the quality of care provided to captive animals (Hosey & Skyner 2007).

Behavioral measurements of wellbeing are complicated by their subjective nature and the variability in baseline behaviors between animals. Genetic susceptibility, age, and physiological state of an animal are just a few aspects that influence a behavioral stress response and its biological cost to the individual (Carstens & Moberg 2000). For example, an animal who is generally at a higher baseline state of arousal may display more obvious responses to an environmental stressor as opposed to an animal with lower levels of arousal, or vice versa. Therefore, the observer must not only be familiar with indicators of distress specific to the species in question, but must also have an established understanding of the personalities of each animal in a colony in order to know which behaviors to flag as concerning (Laudenslager & Boccia 1996). Furthermore, certain behaviors (and specifically stereotypies) may be a result of poor living conditions in the past and are therefore not reflective of current welfare status (Bellanca & Crockett 2002).

**Behavioral Imaging Can Assess Pain and Distress**

Despite the challenges outlined above, there are indeed signs of poor welfare that can be measured in behavior. We can generally surmise that an animal is in a negative state of welfare if drastic changes in its typical behavior are observed (National Research Council 2010). Examples of these changes may include increased or decreased vocalizations, developing stereotyped or self-injurious behaviors, and/or a change in temperament (Tardif et al. 2013). These changes may be subtle, and - critically - no single behavior indicates a definitive change. Moreover, these behavioral changes are only meaningful if compared to the animal's behavioral baseline. As a result, the identification of a large and longitudinal database of that individual animals' behavior must be part of the assessment process. Pose estimation software provides an excellent means to combine the individualization of each animal observed with reduced cost and manpower, as well as improving reliability. Behavioral imaging systems also allow for continuous observations, including outside normal working hours or when staff are unavailable. Constant observation allows more data to be collected on each animal, and presents the opportunity to alert caretakers of immediate threats to the animals' well being on short notice. Ultimately, the most important factor may be one of cost - using humans to assess pain and distress is feasible, if imperfect, but requires expensive highly trained observers.

While behavioral imaging can solve manpower and expense issues, its benefits go beyond that benefit by removing the limitations and biases of humans. By providing a standardized method of assessment, this software removes inter-observer unreliability. It also has the potential to detect gradual changes in the nature and frequencies of behaviors, and subtle signs of pain or discomfort that could go unnoticed by human observers. For example, by assessing abnormal behaviors against the baseline of a specific animal, this software could determine that grooming frequency or intensity has changed over time, allowing caretakers to intervene by removing apparent stressors or by providing enrichment before obvious signs of alopecia occur. Behavioral imaging software may be potentially sensitive enough to detect very subtle behaviors that a trained human observer may not notice, such as a disguised limp. Indeed, a behavioral imaging system could even detect second-order changes in behavior, such as a reduction or delay in walking behavior, that reflect the animal's attempts to hide its pain.

Automated pose tracking systems can also expand our knowledge of animal welfare because they can potentially detect welfare indicators that were previously unknown. This would require novel research - in particular, it would require ground truth data that have been validated and use those to train the system. However, once that is done, those linkages can, if replicated, be used in future welfare efforts. These systems can have a much broader field of view than human perception. For example, deep learning systems can detect behaviors that are too fast or too slow for human observers to detect, that involve multiple small signals, that are too subtle for humans, or that humans may not recognize.

While the issue of pain and distress is perhaps most relevant to the confined spaces and testing requirements of laboratory environments, the same principles hold true in other captive populations, including zoos, wildlife rescues, and field cages. If specialized software can learn to detect pain and distress in lab animals, it has the potential to improve the lives of the animals in any captive setting.

**Behavioral Imaging Can Improve Welfare**

Behavioral imaging has a clear role in identifying pain and distress in animals, but it also has the potential to help actively improve welfare by creating more enriching environments. Historically, attempts to improve welfare of zoo or laboratory housed animals have focused on environmental "inputs" (e.g. cage/enclosure size, addition of enrichment, allowing for social interactions), with less focus on evaluating the animals "output" (i.e. their physiological, health, or behavioral response to these environmental inputs) (Hewson, 2003; Spangenberg & Keeling 2016; Truelove et al., 2020). A focus on environment over physiology, health, or behavior may in part be due to the challenges of measuring these responses (as outlined above). However, despite these challenges, a focus on animal outputs is crucial because efforts to improve welfare based on environmental input may be directed ineffectually or may potentially cause more harm than good. For example, in response to the small enclosure sizes of early zoos, new designs focused on increased space. However, studies have found that quality of space appears to outweigh quantity of space for great apes (Wilson, 1982) and social factors may outweigh spatial factors for macaques (and presumably for several other species as well, Erwin, 1979).

Fortunately, indicators of both optimal and poor welfare manifest as specific behaviors. For example, interventions should increase normal behavior while decreasing self-injurious or otherwise negative behaviors (National Research Council 1998): social companions should lead to an increase in affiliative behaviors rather than be a cause for avoidance and stress; enrichment should encourage exploration and species-typical foraging behaviors rather than produce neophobic responses, conflict with conspecifics, or excessive foraging at the expense of other healthy behaviors. There is a clear need for objective and sensitive measures of behavior, and behavioral imaging provides a new resource to fulfill these aims.

## BEHAVIORAL IMAGING CAN IMPROVE WELFARE WITHIN THE CONTEXT OF THE THREE Rs OF ANIMAL RESEARCH

The three Rs of animal research aim to improve animal welfare by **_Replacement_** of animals with alternative models, **_Reduction_** in the number of animals used for a study, and **_Refinement_** of study methods and housing/husbandry to minimize pain, suffering, and distress (Russell & Burch 1959). As outlined earlier, research on nonhuman primates is both unavoidable and fraught with welfare challenges (Hau & Schapiro 2007). The three Rs therefore represent a compromise in the direction of improving welfare. Behavioral imaging can contribute to each of the three Rs.

In this section, we focus on the benefits of behavioral tracking to neuroscience in particular. We choose neuroscience as an example domain where imaging can contribute to the three Rs partly because of its importance and visibility in primate science, and partly because of our personal interest in the field. Broadly speaking, rhesus macaques (and sometimes Japanese macaques) are often used as a model organism for human brain activity. (As are, increasingly, marmosets, Cyranoski, 2009; Liu et al 2020; Okano & Mitra, 2015; Shimogori, 2018).

### Replacement

The principle of Replacement holds that we can in some cases improve animal welfare by replacing research animals either with "lower" animals (such as rodents) or with animal-free approaches (such as computer models). Classically, neuroscience experiments are bespoke - that means that we devise a hypothesis and design the simplest possible experiment to test it. We are, however, entering into the era of Big Data in neuroscience. That means we can collect data of much higher quantity than in the past. In practice, big data in neuroscience comes from collecting hundreds or even thousands of neurons at a time, collecting whole brain high resolution scans at high field strength, or other methods that provide orders of magnitude more data than traditional bespoke methods. These Big Data methods allow for the collection of so much data that they require new analysis techniques, some of which may not have been invented yet. They also provide the opportunity for post-hoc experimentation - that is, doing experiments on data that already exist in databases.

The major limitation in all of this work, however, is that the understanding of neural data is usually best done in conjunction with behavior. Without behavioral imaging, the behavior will act as a bottleneck - that is, extremely detailed brain data with very simple behavior can only go so far in helping to test neuroscientific hypotheses. On the other hand, the extremely rich data generated by behavioral imaging, when registered with detailed neuronal data, promises to create datasets large enough that they can lead to experiments done in silico for years.

**Reduction**

Reduction refers to the goal of using fewer research animals. Behavioral imaging can also contribute to Reduction, or the design of experiments that use non-human primates in smaller numbers. First, simply by providing more behavior, the need for more animals is correspondingly reduced. That is, the ability to test scientific hypotheses requires a certain amount of data to overcome uncertainties associated with noise. Sometimes, additional data needs to be from independent samples, and in these cases, reduction is impossible (but see below). However, in many cases, the independence of data is not a limiting factor. In these cases, one can get more data with fewer resources.

Second by providing more and richer behavior for each neuron collected, the scientific value of each neuron is enhanced. As a result, the quality of inferences that can be drawn from each neuron is enhanced. Therefore, the scientific community can obtain the same results with fewer research animals.

Third, by providing more naturalistic behavior, the validity of the behavior that is collected is improved. Typically, experiments are performed using unnatural but simple tasks that recapitulate important behaviors, but these tasks may lack external validity. Using more naturalistic tasks, ones with continuous motion and many small decisions, can make the behavior more relevant to answering questions (Yoo et al., 2021). These basic features mean that the number of animals needed to make a given discovery can be reduced. It is not clear yet, because the field is young, how much reduction is possible, but we are sanguine.

To be a bit more speculative, the promise for reduction is potentially even greater. There are some important hypotheses that can be answered with high quality behavioral imaging and without brain activity. For example, we may imagine that some study uses activity in some brain region as an index of some inferred variable. A typical case would be the value of an option to a subject. If the behavior is detailed enough and if enough of the behavior is available to the imaging system, it may be possible to entirely replace the neuronal measure with a behavioral one. This is not to say that imaging will replace invasive measures at all, but in some specific cases it may do precisely that. In other words, the amount of information about the subject's; internal state revealed by behavior may be so great that it obviates the need for invasive measurements.

**Refinement**

Refinement refers to the goal of using techniques that are less invasive or otherwise have negative welfare consequences. NHP behavioral studies commonly require the animal to be placed in a primate chair or have its movement restricted in some other way. While restricting the movements of subjects has practical benefits, there are several scientists beginning to experiment with less restrictive data collection methods such as cage-based cognitive testing in animal housing environments like touchscreen "kiosk" stations (Womelsdorf et al 2021). While these kiosks have scientific benefits, such as improved ecological validity because tasks allow for unrestrained species-typical behaviors, and reduced manpower because the animal does not need to be handled, they also improve welfare by allowing for more free movement in the world, which animals typically find rewarding.

Indeed, in-cage touchscreens have been shown to be a form of cognitive enrichment and also allow for autonomy and provision of choice (e.g. the choice of which task to engage in, at what time, and for how long), which are imperative for psychological well being (Egelkamp et. al. 2016; Egelkamp & Ross 2019). Handling and transfers to primate chairs are also a source of stress eliminated by home-cage kiosks. Touchscreen kiosks also may allow for a Reduction in the number of animals used in the study as a single animal, with access to a kiosk all day, can work multiple times per day, when it is most motivated (an animal's time preferences may change daily and be difficult to predict by investigators). Despite these benefits to research and welfare, there remain unresolved issues with kiosks that may prevent their widespread use. As the animal is housed in a colony room, there will invariably be distractions from conspecifics and husbandry teams, potentially impacting the quality of the data. It is also difficult to determine if the animal is fully engaged in the task (other than trials completed); they may interact more with the task throughout the day because it is available but may be less focused, motivated, or enthusiastic.

Behavioral imaging has the potential to resolve these issues by determining whether trials are completed while the animal is focused and therefore valid or whether, through objective means, data should be discarded. This software can also assist with training. For example, animals can be monitored and rewarded for being calm and focused, and a closed loop system, whereby if an animal deviates from the task or loses engagement there is mitigation or an intervention for re-engagement, can produce cleaner data, refined training, and reduced stress in the animal. Behavioral imaging also has the potential to provide a better understanding of how the animal performs the task. Taken together, by reducing the drawbacks associated with kiosks, behavioral imaging can improve welfare by allowing for increased use of these beneficial devices.

## BEHAVIORAL IMAGING CAN IMPROVE OUR KNOWLEDGE OF "NATURAL" BEHAVIORS AND THIS KNOWLEDGE CAN BE APPLIED TO IMPROVE WELFARE IN CAPTIVE AND WILD POPULATIONS

The field of applied animal behavior focuses on applying our understanding of animal behavior to improve the welfare of captive and wild animals (Fraser & Weary 2021). It often uses the behavior of wild animals to inform this purpose. This goal necessitates both a clear understanding of the frequency and types of behaviors displayed by wild animals in their natural habitats and the differences between these behaviors and those displayed by their captive counterparts, newly released animals, or wild animals in disturbed landscapes. For example, if we determine that wild marmosets spend approximately 10% of their day engaging in predator surveillance, and a group of captive marmosets or wild marmosets living in disturbed landscapes spend 20% of their day engaged in these behaviors, caretakers could add hiding places or canopy cover and reassess time budgets to ensure that predator surveillance, and likely stress, has decreased. On the other hand, we may be able to rest assured that marmosets who spend 10% of their day on surveillance are not a source of concern.

Thus far, research on the behavior of wild animals relies on either direct observation or camera trapping, both of which require large amounts of manpower to identify and code behaviors, and are fraught with issues such as inter-reliability in coding, required time to train coders, and, in the case of direct observation, disruption to natural behaviors due to the presence of a human observer. As a result, while these methods are valuable in identifying broad behavioral categories (e.g. geophagy, predation events, food extraction methods) they are less suited to identifying subtler behaviors and behavioral patterns. The use of behavioral imaging to survey the frequency and types of behaviors displayed by wild animals eliminates many of these issues by providing consistent and unbiased ethograms and can process much larger amounts of data than would be feasible by human coders, in addition to detection of both gross and nuanced behaviors. While pose estimation software cannot eliminate all the shortcomings of camera trapping, many of these remaining issues, such as limited coverage of large territories and disturbances caused by the camera itself, are less important in the context of captivity. For example, while placing camera traps in fixed locations where specific behaviors or resource use occurs biases the types of behaviors observed, these behaviors are most relevant to captive animals; whose enclosures are intended to mimic these important locations.

**Captive Animals**

Designing habitats that allow animals to engage in natural behaviors is of large importance to zoos and other wildlife centers, for both the health and wellbeing of the animals in addition to more rewarding experiences for zoo patrons. Understanding how animals use resources, engage with their space, and interact with each other socially can provide insight into the requirements of a zoo enclosure and allow for continued assessment and improvement of these spaces. Understanding complex social behaviors is also of utmost importance for animals living in restricted spaces with conspecifics and in the case of captive breeding which, depending on the species, can be challenging without a deep understanding of natural courtship and mating behaviors (Swaisgood 2006). Pose estimation software can allow for a better understanding of social organization, signaling behavior, and mate choice, which can inform where resources should be dedicated in order to encourage captive breeding and prevent pair or social group breakdown.

**In Situ Conservation**

With habitat fragmentation and human disturbance ever increasing, it is important to identify the behavioral responses to these threats and determine the best course of action to improve the welfare of these populations. Applied behavioral research is vital to these animal welfare and conservation efforts as behavioral mechanisms play a major role in mediating the impacts of habitat fragmentation and human disturbance on vulnerable species (Yahner & Mahan 1997). Understanding natural behaviors and behavioral responses to new or altered landscapes can inform reserve and habitat corridor design, determine the success of translocated or newly released animals, and identify whether anthropogenic nuisances (e.g. noise, artificial light, roads and traffic, ecotourism) are habituated to or result in chronic stress, altered time budgets, or suppressed reproduction (Swaisgood 2006).

**CONCLUSION**

With non-human primates being such a large focus of laboratory research, zoos, and conservation efforts, there is a clear need to consider the welfare of these highly intelligent, sensitive, and social creatures. While necessary, we currently have poor means to institute such change. Behavioral imaging has the potential to revolutionize welfare management by providing real-time analysis of behavior that can detect signs of pain, distress, or other welfare challenges that would otherwise require unfeasible amounts of manpower and expertise. It can also be used to improve the quality and quantity of data collected in scientific studies, reducing the number of animals required and the pain and stress experienced by those animals, without compromising data integrity. Finally, behavioral imaging can extend beyond mere reduction of pain and suffering to actively improve the lives of captive and wild primates by guiding interventions that allow animals to express the behaviors they would naturally. We are excited to see what this technology holds for the future.

**Funding**

This work was supported by NIH grants R01 MH128177 (to JZ), P30 DA048742 (JZ, BH), R01 MH125377 (BH), NSF 2024581 (JZ, BH) and a UMN AIRP award (JZ, BH) from the Digital Technologies Initiative (JZ), from the Minnesota Institute of Robotics (JZ).

**Author contributions**

All authors contributed to the development and writing of this manuscript.

**Institutional review board statement**

This is a review manuscript and does not include novel research.

**Data availability statement**

This manuscript does not describe data.

**Conflicts of interest**

The authors declare no conflicts of interest.